%% file: main.tex
\documentclass[journal=nalefd,manuscript=article]{achemso}
\usepackage{hyperref}
\usepackage{graphicx}
\mciteErrorOnUnknownfalse

\author{Yves Auad}
\email{yves.maia-auad@universite-paris-saclay.fr}
\author{Cyrille Hamon}
\author{Marcel Tencé}
\affiliation{Laboratoire des Physique des Solides, Université Paris Saclay, CNRS, 91405 Orsay, France}
\author{Hugo Lourenço-Martins}
\affiliation{Max Plank Institute for Biophysical Chemistry, 37077 Göttingen, Germany}
\alsoaffiliation{IV. Physical Institute, University of Göttingen, 37077 Göttingen, Germany}
\author{Vahagn Mkhitaryan}
\affiliation{ICFO-Institut de Ciencies Fotoniques, The Barcelona Institute of Science and Technology, 08860 Castelldefels (Barcelona), Spain}
\author{Odile Stéphan}
\affiliation{Laboratoire des Physique des Solides, Université Paris Saclay, CNRS, 91405 Orsay, France}
\author{F. Javier {Garc\'{\i}a de Abajo}}
\affiliation{ICFO-Institut de Ciencies Fotoniques, The Barcelona Institute of Science and Technology, 08860 Castelldefels (Barcelona), Spain}
\alsoaffiliation{ICREA-Instituci\'o Catalana de Recerca i Estudis Avan\c{c}ats, Passeig Llu\'{\i}s Companys 23, 08010 Barcelona, Spain}
\author{Luiz H. G. Tizei}
\affiliation{Laboratoire des Physique des Solides, Université Paris Saclay, CNRS, 91405 Orsay, France}
\author{Mathieu Kociak}
\email{mathieu.kociak@universite-paris-saclay.fr}
\affiliation{Laboratoire des Physique des Solides, Université Paris Saclay, CNRS, 91405 Orsay, France}

\title[An \textsf{achemso} demo]{Unveiling the coupling of single metallic nanoparticles to whispering-gallery microcavities}


\begin{document}

\begin{abstract}
Whispering-gallery mode resonators host multiple trapped narrowband circulating optical resonances that find applications in quantum electrodynamics, optomechanics, and sensing. However, the spherical symmetry and low field leakage of dielectric microspheres make it difficult to probe their high-quality optical modes using far-field radiation. Even so, local field enhancement from metallic nanoparticles (MNPs) coupled to the resonators can interface the optical far-field and the bounded cavity modes. In this work, we study the interaction between whispering-gallery modes and MNP surface plasmons with nanometric spatial resolution by using electron-beam spectroscopies in a scanning transmission electron microscope. We show that gallery modes are induced over a selective spectral range of the nanoparticle plasmons, and additionally, their polarization can be controlled by the induced dipole moment of the MNP. Our study demonstrates a viable mechanism to effectively excite high-quality-factor whispering-gallery modes and holds potential for applications in optical sensing and light manipulation.
\end{abstract}

\input{fullLetter}
\bibliography{biblio}
\input{04-Supplementary}

\end{document}

%% file: fullLetter.tex
The past few years have witnessed unprecedented advances in photonic technologies. In particular, plasmonics has attained the precise engineering of absorption and scattering properties of metallic nanoparticles due to improvements in synthesis methods \cite{ShapeSizeDependent,SERS,PlasmonEnginnering,PlasmonEnginnering02}. At the same time, fabrication techniques have allowed the widespread use of whispering-gallery mode resonators (WGMRs), which have attracted strong interest in the context of quantum electrodynamics, \cite{OpticalMicroVahala, CavityQED} optomechanics, \cite{OptoMechanics1,OptoMechanics2} and sensing applications \cite{OnChipSpliting,SensingLasingReview}. The tuning capability of both technologies makes the study of coupled systems particularly interesting as we discuss throughout this work.

A physical description of WGMs was first proposed by Lord Rayleigh using acoustic waves circulating the dome of St. Paul's Cathedral \cite{RayleighWGM}, while the electromagnetic-wave analog can be found in circular or spherical microstructures. Resonances are characterized by their transverse electric (TE) or magnetic (TM) polarization and a set of angular, radial, and azimuthal mode numbers ($l$, $q$, and $m$). Besides, WGMRs made from materials with low intrinsic loss, such as silica, can sustain modes that have exceptionally high quality factors \cite{OpticalMicroVahala} (Q factors).

Due to their high Q factor, coupling light to a WGMR is a challenging task. Efficient methods almost always rely on evanescent fields using gratings, prisms, or tapered fibers \cite{PhaseMatchedTaper,QualityWGM,CriticalCouplingTaper,MoysesScattering}. In particular, fiber tapering can reach coupling efficiencies as high as 99\%, but requires active positioning to maintain this level of performance over long time periods. Free-space light is an alternative to evanescent coupling, but it has only encountered partial success with spheres \cite{OpticalResonatorsPartI, OpticalSensing} and asymmetric cavities \cite{ChaoticResonator, DirectionalEmission} having high radiative losses due to the bounded nature of the gallery modes.

A practical solution to the problem is to couple an intermediate (e.g., plasmonic) nanoparticle to the gallery modes and to employ it as an evanescent light coupler into the WGM microresonator \cite{PurcellSiNanocrystals, CavityEnhanced, controlledCoupling, WGMCoatedQD}. This approach has been successfully used to detect and characterize the nanoparticle using mode energy shifting \cite{ProteinDetection,SingleNPDetection,SingleVirusDetection}, splitting, \cite{SplittingHighQ, PurcellModeSplitting} or broadening \cite{SingleDissipativeSensing,BroadeningReview} of the unperturbed resonances. 
Single-metallic-nanoparticle and gallery-mode coupling using far-field light has been studied for applications in photocatalysis, \cite{EngineeringPhotocatalyst} as well as for engineering ultranarrow plasmonic resonances \cite{strongCouplingAuNanorod, FarFieldSecondHarmonic}. Selective coupling into TE and TM modes depending on the incident free-space light polarization has also been observed \cite{FreeSpacePaladium}, finding interesting applications in sensing thanks to the clearer mode identification enabled by this method \cite{SensingLasingReview}. However, a sufficient degree of spatial resolution to study the induced near electric field is still lacking on all of the mentioned studies. 

An alternative to study nanoparticles with high spatial resolution is to use a scanning transmission electron microscope (STEM). Electron energy-loss spectroscopy (EELS) and cathodoluminescence (CL) spectroscopy are techniques that can be performed inside the STEM and can combine the sub-nm probe size with high spectral resolution ($<$30 meV) when monochromatic electron sources are used. EELS and CL have been extensively explored to study small (radius $<$ 150 nm) spherical particles \cite{MappingPlasmonsReview, ClYamamoto, CLStem}, including the observation of low-order optical modes in SiO$_2$ spheres by Hyun \textit{et al}. \cite{HyunWGM}. Large spheres (radius $>$ 1.5 $\mu$m) have been studied with electron beams using photon induced near-field electron microscopy \cite{ZewailGain}, in which a few gallery modes have been observed due to the interaction of an externally applied optical field with the free electrons \cite{OferWGM}. More recently, Müller \textit{et al}. have measured broadband light emission from high-Q WGMRs using fast electrons \cite{Muller2021}, but without the complementary information obtained from EELS. Most of these examples, however, could not resolve a large number of modes. Besides, the bare resonators did not offer much to be spatially explored due to their spherical symmetry.

\begin{figure}[H]
    \centering
    \includegraphics[width=0.57\textwidth]{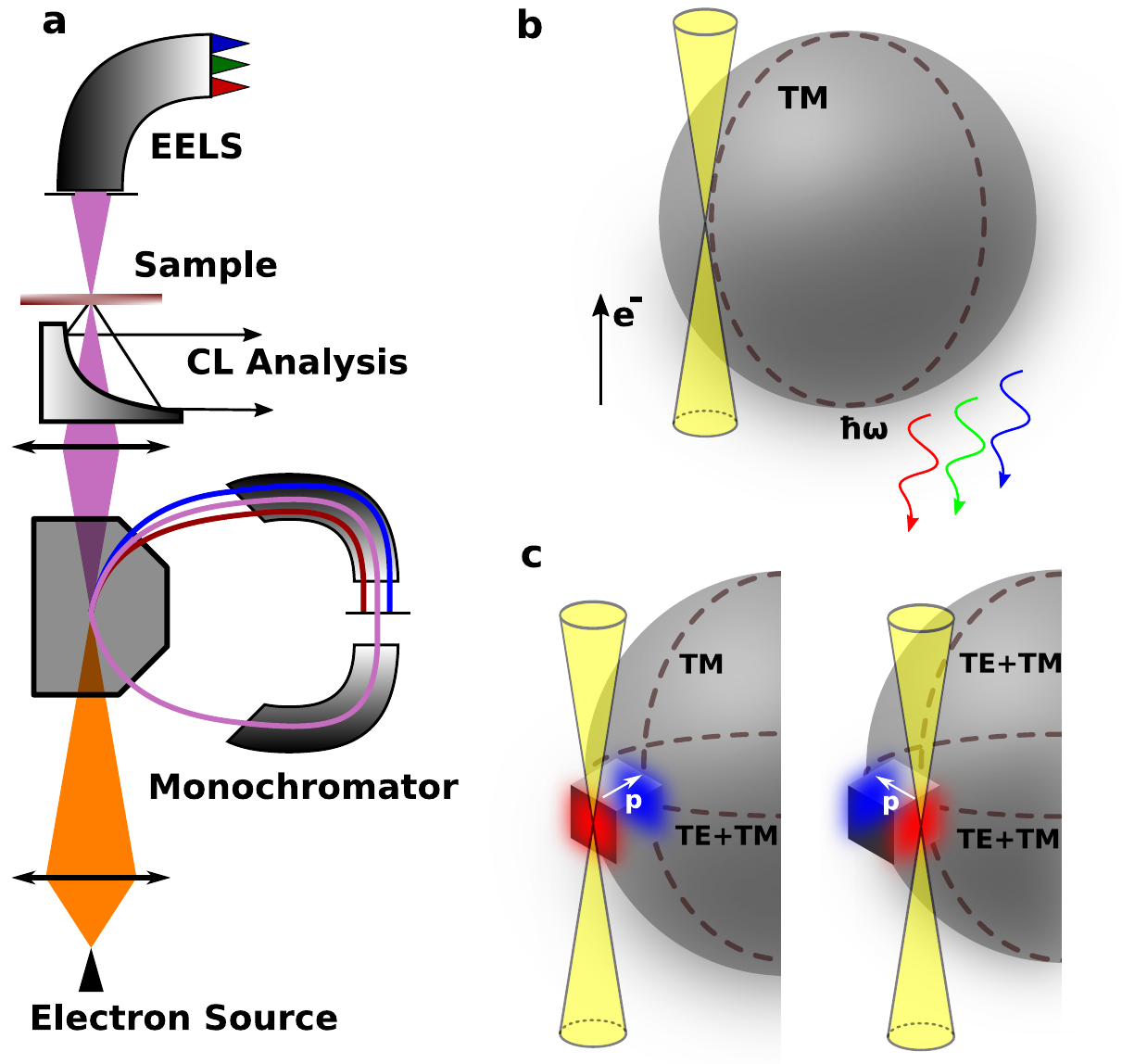}
    \caption{Configuration used to study the coupling between a MNP and a WGMR. (a) Scheme of an electron microscope containing an electron source, an electron monochromator, a CL mirror, the sample and an electron spectrometer. 
    (b) Scheme of an electron beam and the excited TM mode (dashed line) in a bare dielectric sphere. (c) Through the addition of a metallic nanocube, the polarization of the excited WGM can be controlled by the electron-probe position.}
    \label{fig:Intro_Schema}
\end{figure}

In this work, we study the coupling of an electron beam to large SiO$_2$ spheres, in which the narrow-band resonances observed by EELS and CL are attributed to circulating gallery modes in the plane containing the electron trajectory. We explore the coupling of a single silver nanocube with the WGMR through the modulation of the low-Q resonances of the nanocube surface plasmons by the higher-Q gallery modes, as well as by a dependence of the gallery-mode polarization on the orientation of the electron-induced nanoparticle dipole moment, which can be achieved by changing the electron probe position \cite{PlasmonicSilverNanowire, JaysenSingle}. Figure \ref{fig:Intro_Schema} presents a scheme of the experimental setup, as well as illustrations of the influence of the electron beam position on the excitation of modes with different polarizations.

The analytical solution of the energy loss probability of a fast electron upon interaction with a spherical dielectric body can be written as \cite{JavierSingleSphere, JavierClusterDielectrics, JavierReview}

\begin{equation}
    \Gamma_{\rm loss}(\omega) = \frac{1}{c\omega}\sum_{l=1}^{\infty}\sum_{m=-l}^{l}K_m^{2} 
    \left(\frac{\omega b}{v\gamma}\right) \times \\
    \left[C_{lm}^{M} Im(t_l^M) + C_{lm}^{E} Im(t_l^E) \right],
\label{eq:LossProb}
\end{equation}

\noindent
where $K_m$ is a modified Bessel function of order $m$, $b$ is the electron impact parameter with respect to the sphere center, $t_l^M$ and $t_l^E$ are the Mie scattering coefficients \cite{HulstOptics,stratton, MieScatteringOriginal}, which depend exclusively on the sphere radius and its dielectric function, and $C_{lm}^M$ and $C_{lm}^E$ are coupling coefficients that depend on the ratio of electron to light velocities $v/c$. The photon emission probability can be written similarly by making the substitution ${\rm Im}\{t_l^{E, M}\} \rightarrow |t_l^{E, M}|^2$ \cite{JavierSingleSphere}. These simple results show that EELS (CL) is a well-fitted tool to study extinction (scattering) spectroscopy at the nanometer scale, as it is the case in smaller systems\cite{Losquin2015}.

We have used non-functionalized SiO$_2$ spheres from BangsLabs Inc \cite{BangsLabs} ranging from 1.50 $\mu$m to 2.00 $\mu$m in radius for the experiments. We used a modified Nion Hermes 200 fitted with an Attolight Mönch CL system. The monochromatization scheme of the Hermes allows us to reach down to $\sim$ 5 meV spectral resolution and access to energies $>$40 meV at a nominal incident acceleration of 60 kV. This has enabled multiple works in the infrared regime, especially on phonons \cite{Krivanek2014,Hachtel2019,Hage2020} and plasmons \cite{Tizei2020,Mkhitaryan2021}. However, at higher acceleration energies and when focusing on relatively high energy losses and broader energy scales as explored in this work, a spectral resolution of 20 - 30 meV is more appropriate. Experimental results were interpreted by energy-loss simulations done by using a 3D finite-difference time-domain (FDTD) method available in Ansys Lumerical \cite{Lumerical} and based on the work from Cao et al \cite{FDTDElls}. Mie scattering calculations were used to estimate the sphere radius \cite{MieScatteringOriginal, Pymiescatt}. The dispersion values for the SiO$_2$ response were taken from Malitson's work \cite{MalitsonSiO2}.

\begin{figure}[H]
    \centering
    \includegraphics[width=0.7\textwidth]{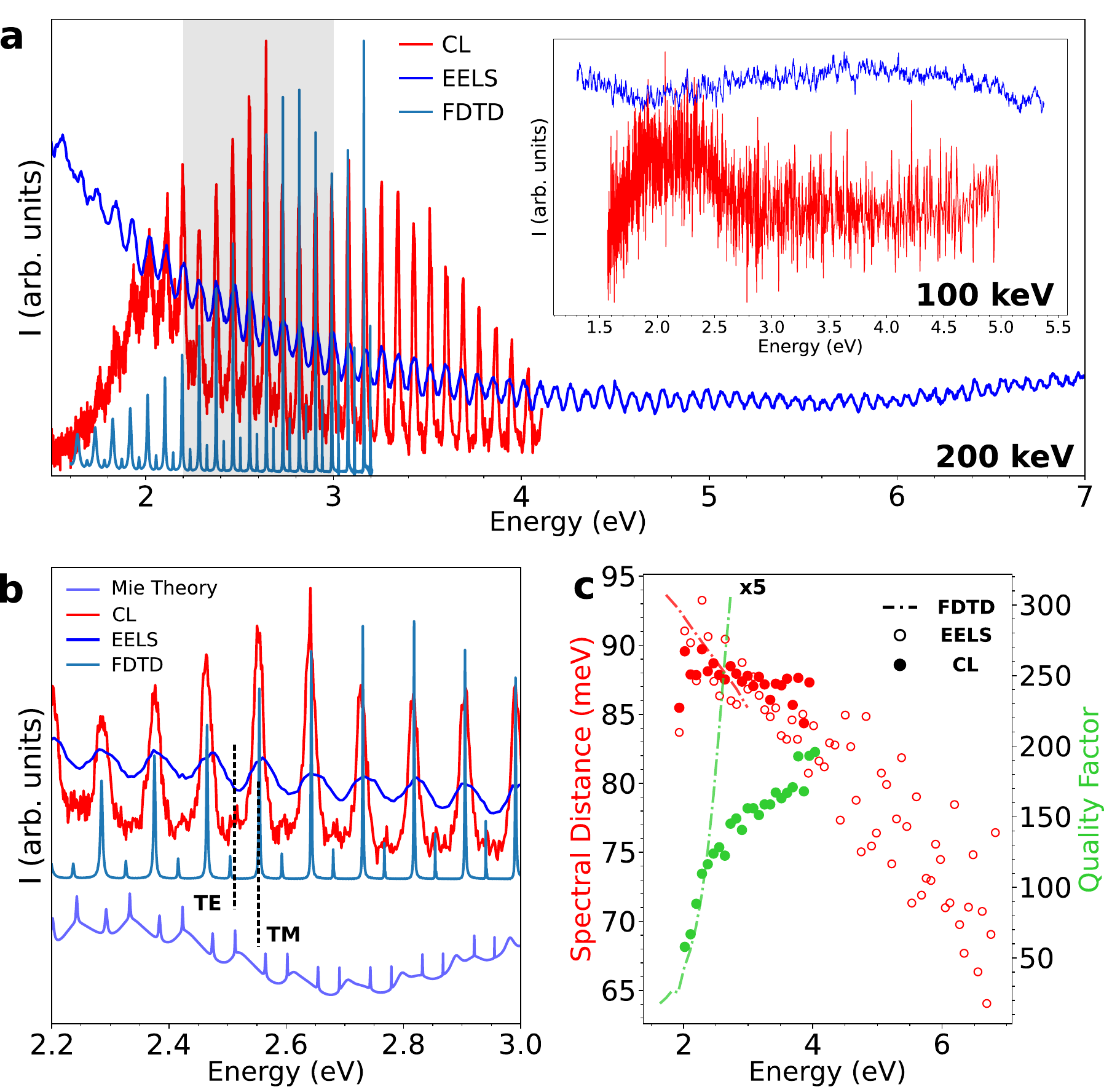}
    \caption{Emergence of optical modes in a SiO$_2$ sphere of 1.595\,$\mu$m radius. (a) EELS and CL spectra measured by using 200 keV electrons, compared to FDTD simulations. The inset shows measurements at 100 keV, lacking any visible resonances. (b) Zoomed area around the highlighted gray rectangle in (a). (c) Measured CL and simulated FDTD Q factors, plotted along with the spectral distance from EELS, CL, and FDTD. The FDTD Q factor is divided by a factor of 5 for readability.}
    \label{fig:PART01}
\end{figure}

Figure \ref{fig:PART01} shows the combined results of EELS and CL using acceleration voltages of 200 kV and 100 kV for one bare sphere suspended on a carbon membrane of $\sim$20 nm thickness. More than 80 resonances were observed in the broadband excitation using 200 keV electrons in EELS, while no energy loss was observed for 100 keV due to the reduced coupling for slower electrons. In both the experimental results and the FDTD simulations, TE polarization is poorly excited by the fast electron, also as a consequence of eq \ref{eq:LossProb} due to the smaller coupling terms of TE in comparison with TM for these electron energies. 

Each resonance in EELS and CL was fitted and the obtained center subtracted from the adjacent mode to form a spectral distance curve, which was subsequently compared with the theoretical free spectral range (FSR)\cite{WGMReview, IdentifyWGMModes}. These values were used for modes from 2.0 eV to 3.0 eV to extract the effective index of refraction. The index was calculated to be 1.398 at 2.0 eV and 1.409 at 3.0 eV, which are values that agree within $\approx$ 95\% with reference values \cite{MalitsonSiO2}. Beyond 3.0 eV, modes have Q factors exceeding $10^{5}$ from Mie Theory, which would be easily washed out by losses and finite energy resolution of our experimental setup. This is specially important for the far-UV resonances observed in EELS, as multiple sets of radial orders combined with the limited spectral resolution of the electron beam undermine mode order identification and the estimative of the index of refraction (see Supplementary Material (SM) for further details). The Q factors from FDTD follow an exponential-like profile as radiation leakage is the only source of loss in the simulation. The inflexion point near 2.8 eV in the experimental Q factor from CL can be attributed to a combination of experimentally induced losses, such as the effect of the carbon membrane and surface inhomogeneities, \cite{WGMReview, QRadMatCompleteMicroSpheres} as well as the expected quality factor reduction from the increased radial order.

\begin{figure}[H]
    \centering
    \includegraphics[width=1.0\textwidth]{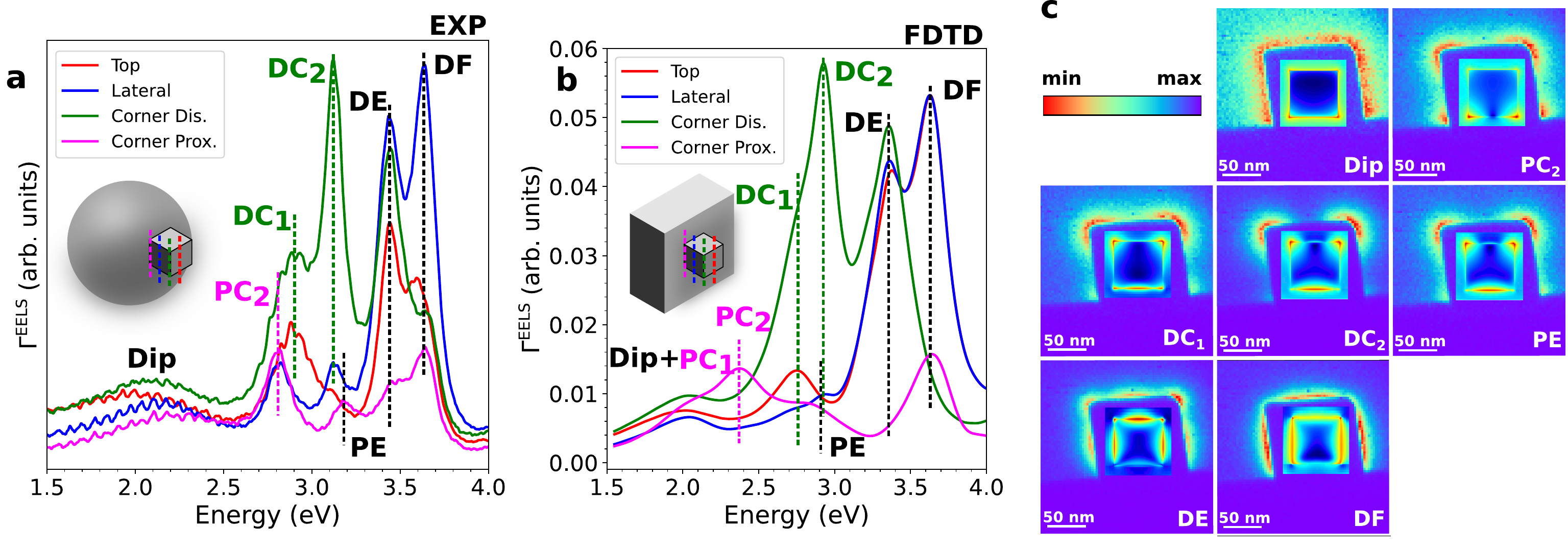}
    \caption{Characterization of MNP-WGMR coupling. (a) Experimental electron energy-loss spectra of a silver nanocube of $\approx105\,$nm side length placed on a silica sphere for four different probe positions as indicated by the inset scheme. (b) FDTD simulated electron energy-loss spectra of a silver nanocube of 100\,nm on a SiO$_2$ planar substrate for similar probe positions. (c) Spectral maps obtained by EELS for all identified modes and, superimposed, the FDTD-simulated absolute value of the electric field.}
    \label{fig:PART02}
\end{figure}

To study the coupling of nanoparticles with WGMRs, we have drop-casted silver nanocubes of $\approx$ 100 - 120 nm side length, synthesized by seed mediated growth, \cite{Zheng2014, Mayer2015} in the sample grid containing the SiO$_2$ spheres. To mitigate the direct coupling of the electron beam with the WGMR, the electron acceleration is kept at 100 keV, where the coupling terms for angular modes $l\approx20$ are reduced by two orders of magnitude (see SM). In theory (for a lossless Drude Metal), bare nanocubes are known to support an infinity number of optical modes \cite{Fuchs1975, CubeInfinity}, which are conventionally divided into corner (C), edge (E), and face (F) modes \cite{Cube2005, UltraLocalCubes,Li2021, Lourenco-Martins2017}. 
The presence of the SiO$_2$ sphere, as a substrate, induces mode hybridization for each of the C, E, and F modes, leading to the so-called proximal and distal splitting in reference to the induced fields concentrated close or opposite to the substrate \cite{CubeHybrid, 3DSurfacePlasmon}, respectively. Figure \ref{fig:PART02} shows the most notable of these modes identified simultaneously by EELS and FDTD simulations, done by placing a 100 nm Ag nanocube on a 500 nm thick SiO$_2$ plane substrate in order to observe the cube-substrate hybridization. Weak coupling between the gallery modes circulating in the sphere and the cube surface plasmons is observed in the spectral range of the dipolar mode (Dip) and, to a lower extent, in the first observed distal corner (DC$_1$) resonance, as shown in Figure \ref{fig:PART02}a.

Loss spectra were studied with the electron probe positioned at the cube top (opposite edge from the substrate), side (lateral edge), distal corner (opposite corner from substrate), and proximal corner (closer corner to substrate) positions, which contain all possible symmetries of the problem. No gallery modes were observed in the proximal corner mode PC$_2$, the second distal corner DC$_2$, the proximal and distal edge modes (PE and DE, respectively), and the distal face mode DF, even though similar bare resonators exhibited detectable modes up to 7 eV under 200 keV electron excitation. This behavior can be attributed to the near-zero net dipole moment of these higher-order modes, and thus, the resulting electric field is not enough to be observed in the EELS spectrum. Figure \ref{fig:PART02}c shows filtered spectral maps of the cube modes, which match the already-known tomographic reconstructions of this system \cite{3DSurfacePlasmon} and were also observed in the FDTD simulations. The absolute value of the electric field obtained from FDTD for each of the cube modes is shown in the respective experimental cube map. Finally, we have performed FDTD simulations of an entire spectral image that confirm the observed spatial and spectral features of the nanocube (see SM).

\begin{figure}[H]
    \centering
    \includegraphics[width=0.7\textwidth]{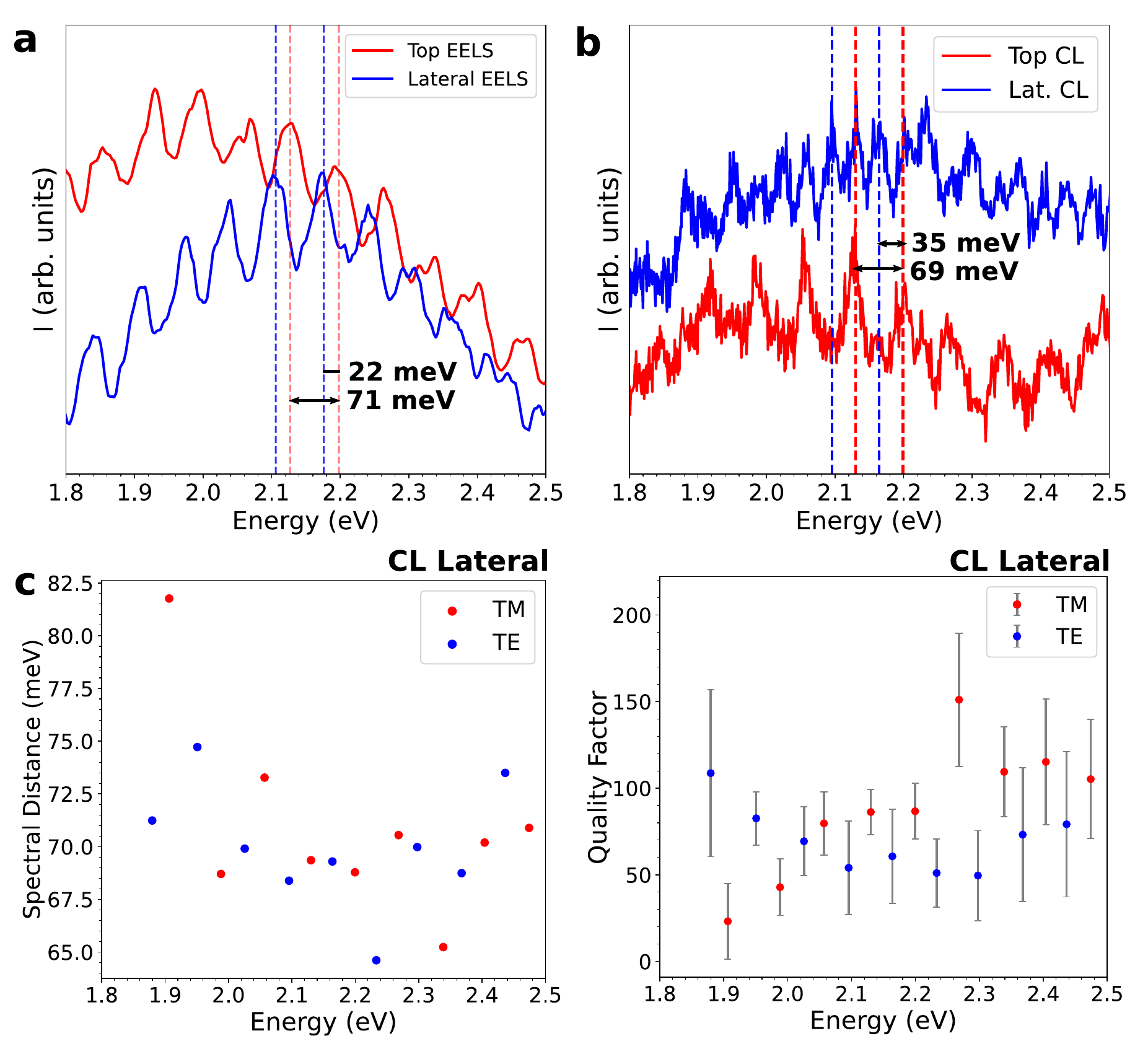}
    \caption{Polarization dependence of the NMP-WGMR coupling. (a,b) Polarization dependence on the probe position as observed in the EELS (a) and CL (b) signals associated with the dipolar mode. TM and TE modes are simultaneously observed in CL for the lateral probe position, while they are unobservable in EELS due to the limited spectral resolution of the electron source. (c) Spectral distance and quality factors for the resonances found with the lateral probe position. The uncertainty in the spectral distance is $<1\,$meV for all data points.}
    \label{fig:PART02_POL}
\end{figure}

Due to the much more complex distribution of the local electric field associated with these higher-order modes, we have focused our analysis on the dipolar mode (Dip) of the cube, which is characterized by dipole moments along the three main symmetry directions. Within the dipolar picture, we suggest that different probe positions induce different gallery-mode polarizations that translate into net dipolar moment directions. As TE modes have no radial electric fields ($\textbf{r} \cdot \textbf{E} = 0$), the top probe position primarily excites TM modes. If the probe is placed laterally to the cube, the resulting electric field is mostly tangential to the sphere surface and can thus excite both TE and TM modes \cite{MethodPredictWGM, FreeSpacePaladium}. The experimental results from EELS show resonances that are equally spaced by $\sim$ 71 meV for each top and lateral probe positions, but shifted $\sim$ 22 meV between each other, as shown in Figure \ref{fig:PART02_POL}. This value is smaller to the fwhm derived from the EELS spectra and therefore limits the resolution of possible neighboring TE and TM resonances. CL offers a simple solution to further explore the problem thanks to its superior spectral resolution. 

The employed CL system is angle-selective and can only detect gallery modes that circulate close to the plane containing the electron trajectory due to the mirror position relative to the sample and the electron beam direction. Experimental CL results are shown in Figure \ref{fig:PART02_POL}b. For the top probe position, which induces a strong radial electric field, TM polarization is more clearly resolved, while for a lateral probe both sets of TM and TE resonances can be observed. The measured FSR is $\sim$ 69 meV and the TE-TM spectral distance can be directly determined with the lateral probe spectrum as $\sim$ 35 meV.

We have fitted the resonances measured in the lateral probe position and divided them into two different sets. Consecutive modes were subtracted to obtain a spectral distance curve and the standard deviation was used to estimate Q factors. We have also included error bars for the Q factors due to the non-negligible fitting uncertainties. The spectral distance curve shows no notable difference between polarizations, as expected from its definition. The Q factor for TE modes has a minimum value centered in the cube dipole resonance, which is expected because cube-induced losses are then maximal. For the TM modes, we observed a steady Q factor decrease down to the lowest energy resonance at $\sim$ 1.9 eV, but not a clear minimum. Unfortunately, the low emitted light intensity and the impossibility of easily changing the electron acceleration preclude further exploration of the coupled system. 

It is worth mentioning that CL results outside the dipolar energy range were also obtained. While induced gallery modes were not observed in EELS for DC$_2$ (Figure \ref{fig:PART02}a), CL clearly displays a coupling component (Figure 
S5
in SM). This observation reinforces the argument that the coupling strength is dependent on the net dipole moment. Since CL measures the radiative component of the coupled plasmon-WGM mode, induced gallery modes can be observed. In addition, EELS is a combination of radiative and non-radiative losses and, because the non-radiative contribution only contains plasmonic losses and dominates over the radiative component, gallery modes are difficult to resolve in the EELS spectra.

To obtain further insight into the physics of the nanocube-sphere system, we have developed a semi-analytical model that captures the main elements of the experiment. In this model, the MNP is described as a dipolar scatterer of polarizability $\alpha(\omega)$ whose dielectric environment is modified by the presence of the dielectric sphere. The latter enters through a $3\times3$ Green tensor $\mathcal{G}$ that transforms the effective particle polarizability into $\alpha^{\rm eff}=1/(\alpha^{-1}-\mathcal{G})$, and whose components are obtained from the electric field induced at the position of a unit dipole placed at the nanocube center. We calculate $\alpha(\omega)$ from a finite-difference method (see Figure\ 
S6
in SM) and $\mathcal{G}$ using the boundary-element method \cite{JavierReview} in the presence of the sphere. 

\begin{figure}[H]
    \centering
    \includegraphics[width=0.7\textwidth]{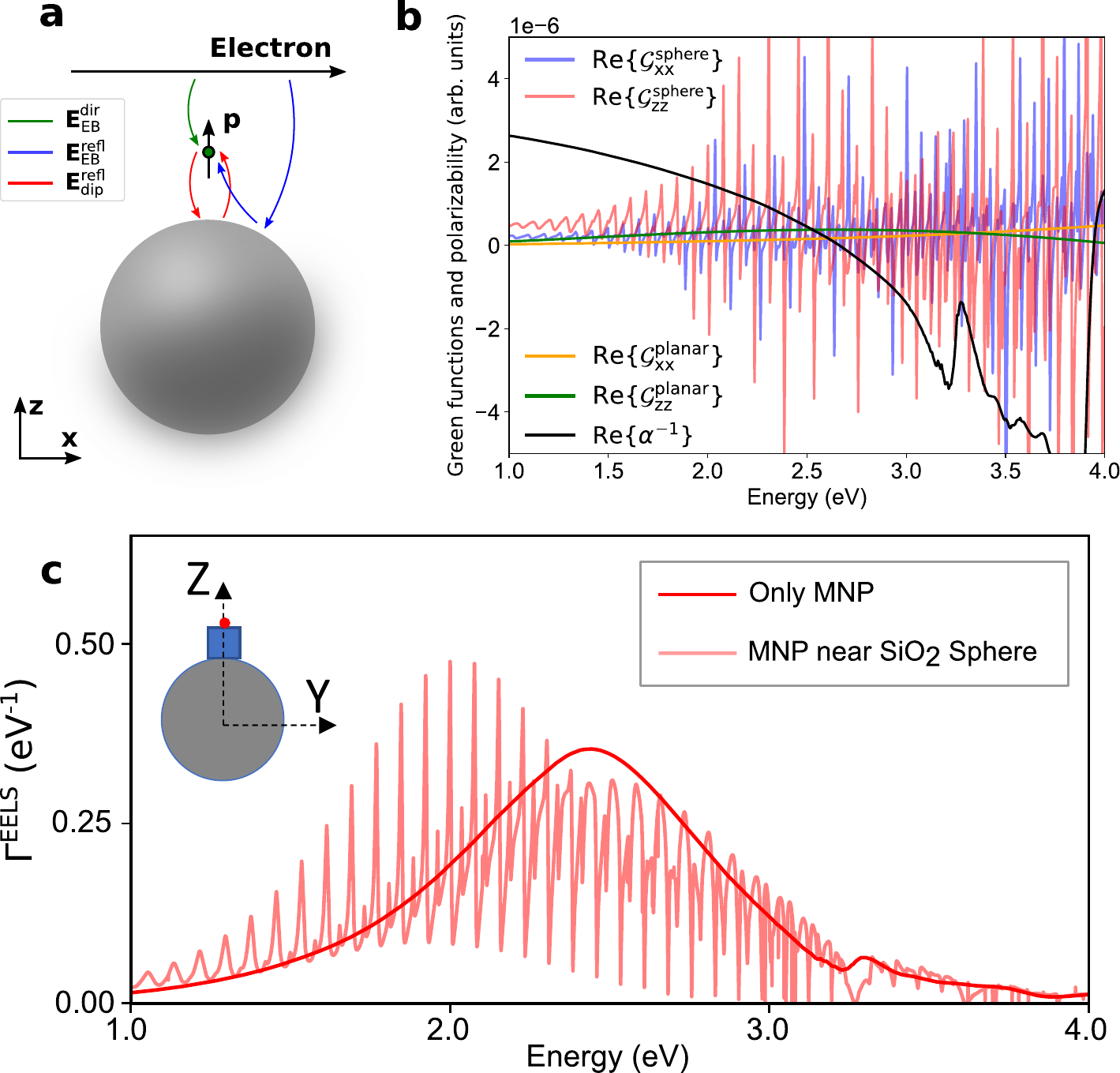}
    \caption{Analytical description of free-electron interaction with a MNP-WGMR hybrid system. (a) Electric field contributions to the self-consistent dipole model (see main text), where the MNP is treated as a dipolar scatterer. (b) Calculated electromagnetic Green tensor components for the self-induced field produced by the particle dipole on itself due to the presence of the dielectric sphere, compared with the real part of the inverse of the MNP polarizability. (c) Model calculation of the EELS probability corresponding to the nanocube alone and the nanocube placed near the sphere for the electron beam position shown in the inset.}
    \label{fig:THEO_SELF_CONSISTENT}
\end{figure}

These elements are represented in Figure\ \ref{fig:THEO_SELF_CONSISTENT}b, where ${\rm Re}\{\alpha^{-1}\}$ is found to change its sign around 2.5 eV, indicating the emergence of a prominent particle plasmon, whereas $\mathcal{G}$ displays sharp oscillations revealing the effect of coupling to the Mie modes of the sphere. For comparison, we show $\mathcal{G}$ for a planar silica surface, which shows a featureless profile. From these considerations, we understand that we are in the weak-coupling regime because the lifetime of the MNP dipole is much smaller than that of the whispering-gallery modes, so that the imaginary part in the denominator of $\alpha^{\rm eff}$ remains relatively large and dominated by the nanocube component. Then, the optical response of the MNP-sphere hybrid system is enhanced at the points in which the real part of the denominator is cancelled, as indicated by the crossings between ${\rm Re}\{\alpha^{-1}\}$ and ${\rm Re}\{\mathcal{G}\}$ in Figure\ \ref{fig:THEO_SELF_CONSISTENT}b. The effective dipole induced on the particle receives contributions associated with different scattering paths, as schematically shown in Figure\ \ref{fig:THEO_SELF_CONSISTENT}a. Namely, it is contributed by the direct field produced by the electron ($\textbf{E}_{\rm EB}^{\rm dir}$), as well as by the scattering of this field at the sphere ($\textbf{E}_{\rm EB}^{\rm refl}$) and the scattering of the dipole field at the sphere acting back on the dipole ($\textbf{E}_{\rm dip}^{\rm refl}$); these contributions are all captured in $\alpha^{\rm eff}$, from which a scattering series can directly be constructed by Taylor expanding it in powers of $\alpha \mathcal{G}$.

To calculate EELS in this analytical model, we use the multiple-elastic-scattering of multipole expansions (MESME) method \cite{paper025} with the sphere and the MNP acting as scattering centers. In particular, the sphere is described by multipoles up to order $>30$ and the MNP through the electric dipolar components of the scattering matrix. This method captures all scattering paths schematically represented in Figure\ \ref{fig:THEO_SELF_CONSISTENT}a. The results presented in Figure\ \ref{fig:THEO_SELF_CONSISTENT}c confirm the analysis based on $\alpha^{\rm eff}$: the EELS probability with the MNP alone exhibits a prominent plasmon, but it is modulated through coupling to Mie resonances of the sphere in a way similar to that observed in experiment. Note that the model successfully describes the sphere and MNP coupling within the dipolar picture, but it does not account for higher-order plasmon modes. In this sense, although Figure \ref{fig:THEO_SELF_CONSISTENT} shows gallery modes along the entire displayed energy range, they are not observed in Figure \ref{fig:PART02}a because higher-order modes mask the tail of the dipolar resonance. Further discussion of the analytical model can be found in the SM.

In conclusion, we have described, with high spatial resolution, the coupling of MNPs to WGMRs using fast electrons, observed from both energy absorption (EELS) and light emission (CL) measurements. While CL can be used to improve the experiment spectral resolution, EELS provides rich and complete absorption information over a large spectral range: a compelling example on how EELS and CL can be used together to provide self-complementary information. The combined EELS-CL measurements would be even more relevant if the electron energy could be changed without causing major microscope misalignment. Beyond its remarkable agreement with experiments, FDTD-simulated energy-loss spectroscopy provides us with a deep physical insight into the coupling mechanism. Finally, the plasmon resonance excitation by the electron probe allows for the manipulation of the gallery-mode polarization by either coupling to TM or to both TM and TE resonances. With the advent of a new generation of monochromated STEMs, experiments requiring such high spectral and spatial resolution are now possible. With this work, we help to open the path for the high quality-factor and small modal volume photonic devices to be scrutinized in a STEM.

\section{Supporting Information}
The supporting information contains details on the electron coupling probability for bare WGMRs; fitting results for bare SiO$_2$ spheres; FDTD simulations for the hybridization of cube modes on a flat SiO$_2$ substrate; CL results for non-dipolar modes in the WGM-plasmon coupled system; theoretical details of the metallic nanoparticle polarizability; and additional results from the analytical model.

\begin{acknowledgement}
The present project has been funded partially by the National Agency for Research under the program of future investment TEMPOS-CHROMATEM (reference no. ANR-10-EQPX-50), from  the  European  Union’s  Horizon  2020  research  and  innovation  programme  undergrant  agreement  No  823717  (ESTEEM3)  and  101017720  (EBEAM) and from the French National Research Agency (QUENOT ANR-20-CE30-0033). VM and FJGA acknowledge support from European Research Council (Advanced Grant No. 789104eNANO), Spanish MCINN (PID2020-112625GB-I00 and CEX2019-000910-S), Generalitat de Catalunya (CERCA and AGAUR), and Fundaciós Cellex and Mir-Puig.
\end{acknowledgement}

\section{Notes}
MK patented and licensed technologies concerning the Attolight Mönch CL system and is a part-time consultant for the company. All other authors declare no competing financial interests.

%% file: 04-Supplementary.tex
\setcounter{figure}{0}
\makeatletter 
\renewcommand{\thefigure}{S\@arabic\c@figure}
\makeatother

\clearpage
\begin{normalsize}
\textbf{Supplementary Material for Unveiling the coupling of single metallic nanoparticles to whispering-gallery microcavities}
\end{normalsize}

\begin{center}
\author{Yves Auad$^1$, Cyrille Hamon$^1$, Marcel Tencé$^1$, Hugo Lourenço-Martins$^{2, 3}$, Vahagn Mkhitaryan$^4$, Odile Stéphan$^1$, F. Javier {Garc\'{\i}a de Abajo}$^{4, 5}$, Luiz H. G. Tizei$^1$, Mathieu Kociak$^1$}

\vspace{5mm}

\date{%
    $^1$Laboratoire des Physique des Solides, Université Paris Saclay, 91405 Orsay, France\\
    $^2$ Max Plank Institute for Biophysical Chemistry, 37077 Göttingen, Germany\\
    $^3$ IV. Physical Institute, University of Göttingen, 37077 Göttingen, Germany\\
    $^4$ICFO-Institut de Ciencies Fotoniques, The Barcelona Institute of Science and Technology, 08860 Castelldefels (Barcelona), Spain \\
    $^5$ICREA-Instituci\'o Catalana de Recerca i Estudis Avan\c{c}ats, Passeig Llu\'{\i}s Companys 23, 08010 Barcelona, Spain
}
\end{center}

\subsection{A - Electron coupling probability for bare WGMRs}
As discussed in the main text, eq 
1
provides an analytical solution for the EELS probability of a fast electron interacting with a spherical particle. While the Mie coefficients ($t_l^M$ and $t_l^E$) depend exclusively on the sphere radius and its dielectric function, the coupling coefficients ($C_{lm}^M$ and $C_{lm}^E$) depend on the ratio of the electron and light speeds $v/c$. Inspection of the coupling terms shows that, for small spheres, only low angular momentum orders contribute significantly to the energy loss. Moreover, the contribution from magnetic modes (TE) is only comparable to that from electric modes (TM) when $v$ approaches $c$ \cite{JavierSingleSphere, JavierClusterDielectrics}. Additionally, for lossless materials (i.e., with real dielectric functions), Mie scattering terms satisfy ${\rm Im}\{t_l^{E, M}\} = |t_l^{E, M}|^2$ \cite{MieDerivatives} and hence eq 
1
predicts identical results for EELS and CL.

Although intricate, the analytical solution for an arbitrary number of angular modes for a single sphere is possible computationally \cite{JavierSingleSphere}. A convenient way of studying the electron velocity dependence in EELS is by performing finite-difference-time-domain (FDTD) simulations. To calculate the energy-loss probability, one simply applies the following equation to the obtained induced electric field:

\begin{equation}
    \Gamma_{EELS}(\omega) = \frac{e^2}{4 \pi \omega^2} \int \int_{+\infty}^{+\infty} cos \left[(z - z')\frac{\omega}{v}\right] \times Im\left[ \frac{E^{ind}_{z}(z, \omega)}{p(z', \omega)}\right]dz dz',
    \label{eq:SUP_EELS}
\end{equation}

\noindent
where $z$ is the position of the calculated field, $z'$ is the dipole charge position, and $p(z', \omega)$ is the amplitude of the dipole used to mimic the $\omega$ components of the electron as an external source. Simulations were performed using the Ansys Lumerical software \cite{Lumerical}. Further details on how to perform FDTD for EELS can be found in the work of Cao et al \cite{FDTDElls}. Note that the electron velocity only appears on the cosine term, independent from the induced fields along the electron trajectory, which means that the EELS probability for any electron velocity can be calculated by post-processing the same induced fields.

\begin{figure}[H]
    \centering
    \includegraphics[width=0.95\textwidth]{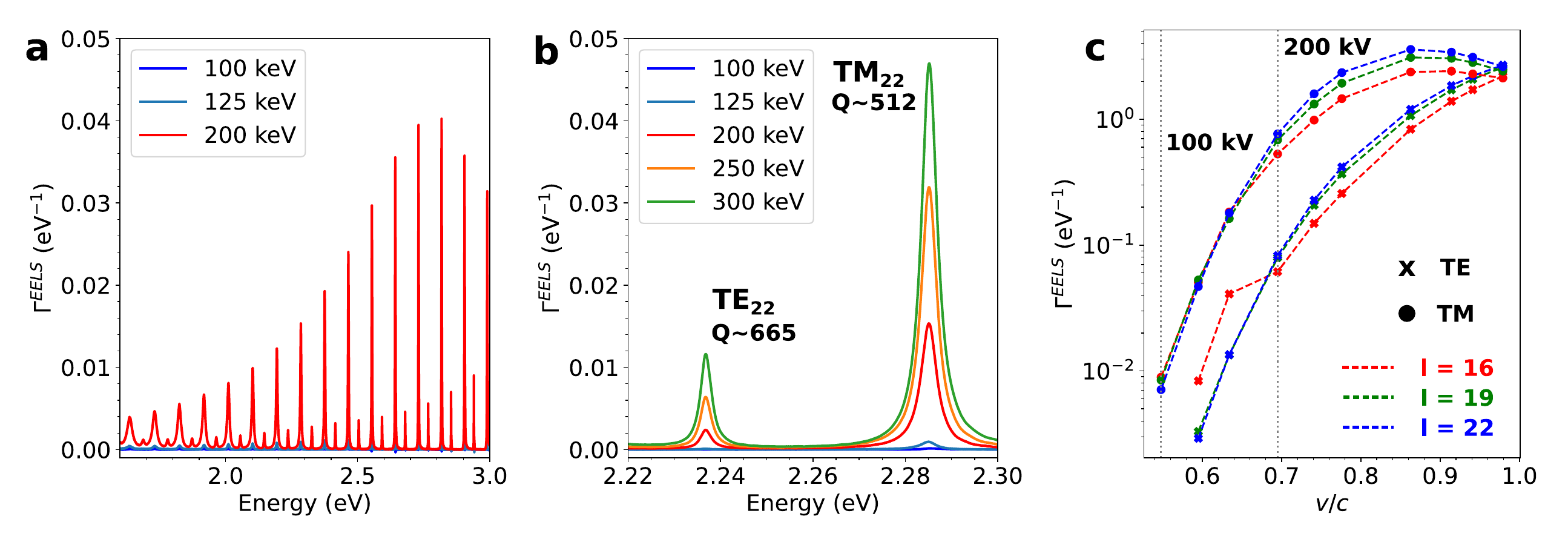}
    \caption{FDTD simulation results for a bare SiO$_2$ sphere. (a) EELS probability over a broadband energy range for 100 keV, 150 keV, and 200 keV electrons and a sphere of radius $\approx$ 1.60 $\mu$m. (b) Details of the TM$_{22}$ and TE$_{22}$ resonances for multiple electron kinetic energies. (c) The probability for three TM and TE resonances as a function of $v/c$.}
    \label{fig:SUP_WGRM_EE}
\end{figure}

The EELS probability for a bare resonator was calculated for several $v/c$ ratios, as shown in Figure \ref{fig:SUP_WGRM_EE}. We have inspected a few angular order peaks close to the visible range for a sphere of $\approx$ 1.60 $\mu$m. For the TM$_{22}$ mode, for example, the EELS probability increases approximately two orders of magnitude as the electron energy is increased from 100 keV to 200 keV. Also, the loss probability for the TE modes approaches the TM ones only when $v \sim c$, as expected. Finally, note that the quality factor for the TM polarization is slightly smaller than the TE one, which is an expected result when only radiation losses are taken into account \cite{WGMReview}.

\clearpage
\subsection{B - Fitting results for SiO$_2$ bare spheres}

\begin{figure}[H]
    \centering
    \includegraphics[width=0.9\textwidth]{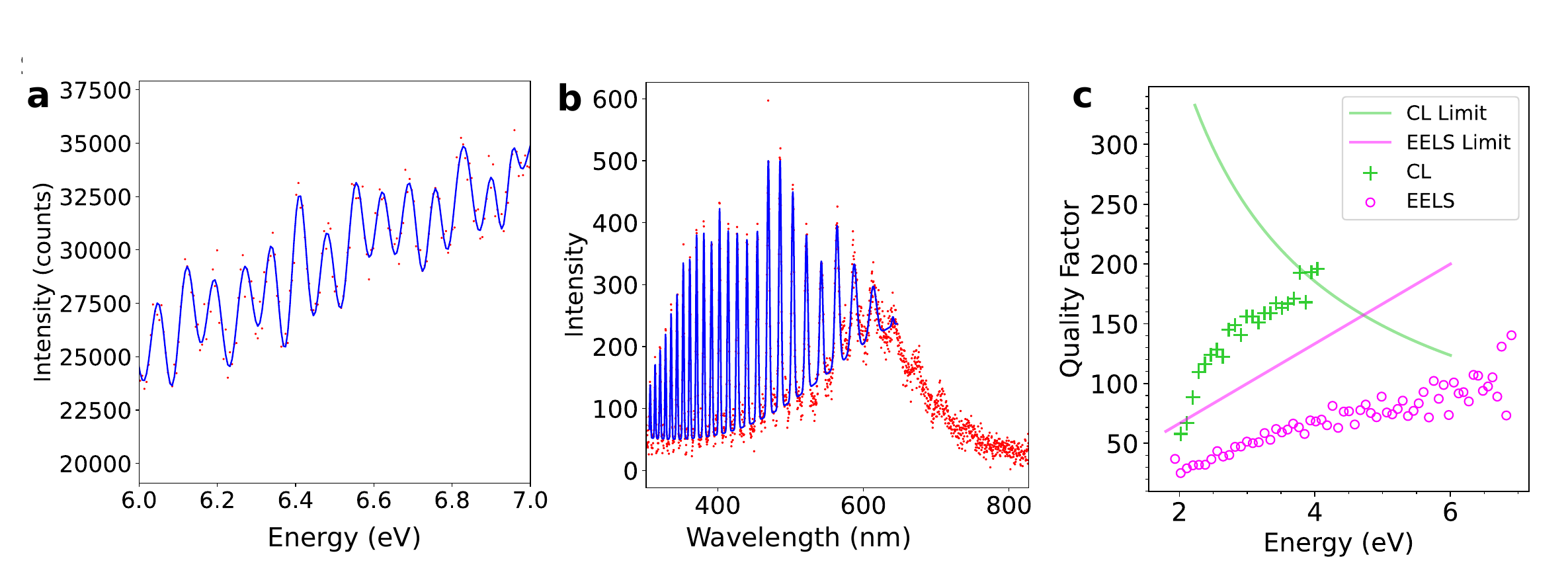}
    \caption{Fitting results for EELS and CL. (a) Fitting of EELS peaks obtained at 200 keV. (b) Fitting results for CL. (c) Quality factor obtained from both techniques as a function of energy, including the theoretical limit of the maximum detectable Q-factor.}
    \label{fig:SUP_CL_FITTING}
\end{figure}

Due to the large amount of spectrally close resonances in bare resonators for EELS / CL, here we describe how the fitting procedure was performed, along with the limitations from both types of measurements. All data sets were fitted using Gaussians for each observed resonance and a power law for the background signal. Each Gaussian was allowed to shift 2 nm for CL and 30 meV for EELS from the initial guess, and standard deviations were free to be determined as long as they remained less than 15 nm wide in CL and 100 meV in EELS. Data modeling was entirely carried out in Hyperspy \cite{Hyperspy}. The quality factor was calculated as $Q=E/\Delta E$, where $\Delta E$ is the resonance FWHM, determined from the Gaussian standard deviation as $\Delta E = 2 \sqrt{2 \ln{2}} \sigma \approx 2.355 \sigma$. The Q-factor presented in the main text was obtained for CL measurements due to its improved spectral resolution with respect to EELS and because only the last few points started to approach our instrument point spread function (PSF), determined as $\approx$ 0.7 nm. The PSF limits the maximum observable standard deviation, and thus, if $Q^{CL}_{lim} \approx \lambda / \Delta \lambda_{PSF} \approx 1239.8 / (E \Delta \lambda_{PSF})$, the maximum observable quality factor is inversely proportional to the photon energy. For EELS, in contrast, the PSF can be roughly approximated by the FWHM of the zero-loss peak, and likewise, $Q^{EELS}_{lim}  \approx E / \Delta E_{PSF}$, which has a linear dependence on the photon energy. Figure \ref{fig:SUP_CL_FITTING}c shows the observed quality factors for both techniques, along with the estimate Q-factor limit as a function of energy. The clear linear dependence of the Q-factor as a function of energy for EELS suggests that the measurements are indeed limited by the electron spectrometer resolution. The fact that the experimental curve is below the EELS limit curve is probably due to deviations of the actual system from the model assumptions, since the zero-loss peak is not a perfect Gaussian and the spectral distance between resonances ($\sim$ 80 meV) is comparable to the zero-loss FWHM. Finally, Table \ref{tab:SUP_CL} details the fitting results for CL at 200 keV.

\begin{table}[ht]
\centering
\begin{tabular}{||c c c c || c c c c ||}
\hline
I (cts) & Center (nm) & $\sigma$ (nm) & Q & I (cts) & Center (nm) & $\sigma$ (nm) & Q\\
\hline

3452.5 & 613.42 & 4.50 & 57.87 
& 3062.4 & 402.57 & 1.10 & 156.06 \\

4190.9 & 587.39 & 3.73 & 66.8
& 2634.2 & 391.5 & 1.10 & 151.23 \\

4652.5 & 563.92 & 2.71 & 88.38 
& 2562.1 & 380.95 & 1.02 & 158.74 \\

3002.5 & 542.26 & 2.10 & 109.66
& 2490.4 & 371.02 & 0.99 & 158.71 \\

3620.5 & 521.79 & 1.90 & 116.41 
& 2044.4 & 361.71 & 0.92 & 167.15 \\

4418.9 & 503.13 & 1.72 & 123.93
& 1984.5 & 352.73 & 0.92 & 163.14 \\

4926.6 & 485.65 & 1.61 & 128.40
& 1554.0 & 344.20 & 0.88 & 166.89 \\

5146.7 & 469.50 & 1.63 & 122.47
& 1287.1 & 336.04 & 0.83 & 170.90 \\

3141.1 & 454.44 & 1.38 & 145.11 
&924.1 & 328.41 & 0.72 & 192.77 \\

2900.0 & 440.16 & 1.26 & 148.79 
& 888.9 & 320.96 & 0.81 & 167.98 \\

3143.1 & 426.84 & 1.29 & 140.66  
& 634.5 & 314.10 & 0.69 & 193.19 \\

2829.2 & 414.38 & 1.13 & 156.01
& 438.8 & 307.31 & 0.67 & 195.87 \\

\hline
\end{tabular}
\caption{Intensity, center, and sigma values for each Gaussian fitting of CL at 200 keV.\label{tab:SUP_CL}}
\end{table}

\clearpage
\subsection{C - FDTD-simulated cube mode hybridization on a flat SiO$_2$ substrate}

\begin{figure}[H]
    \centering
    \includegraphics[width=0.8\textwidth]{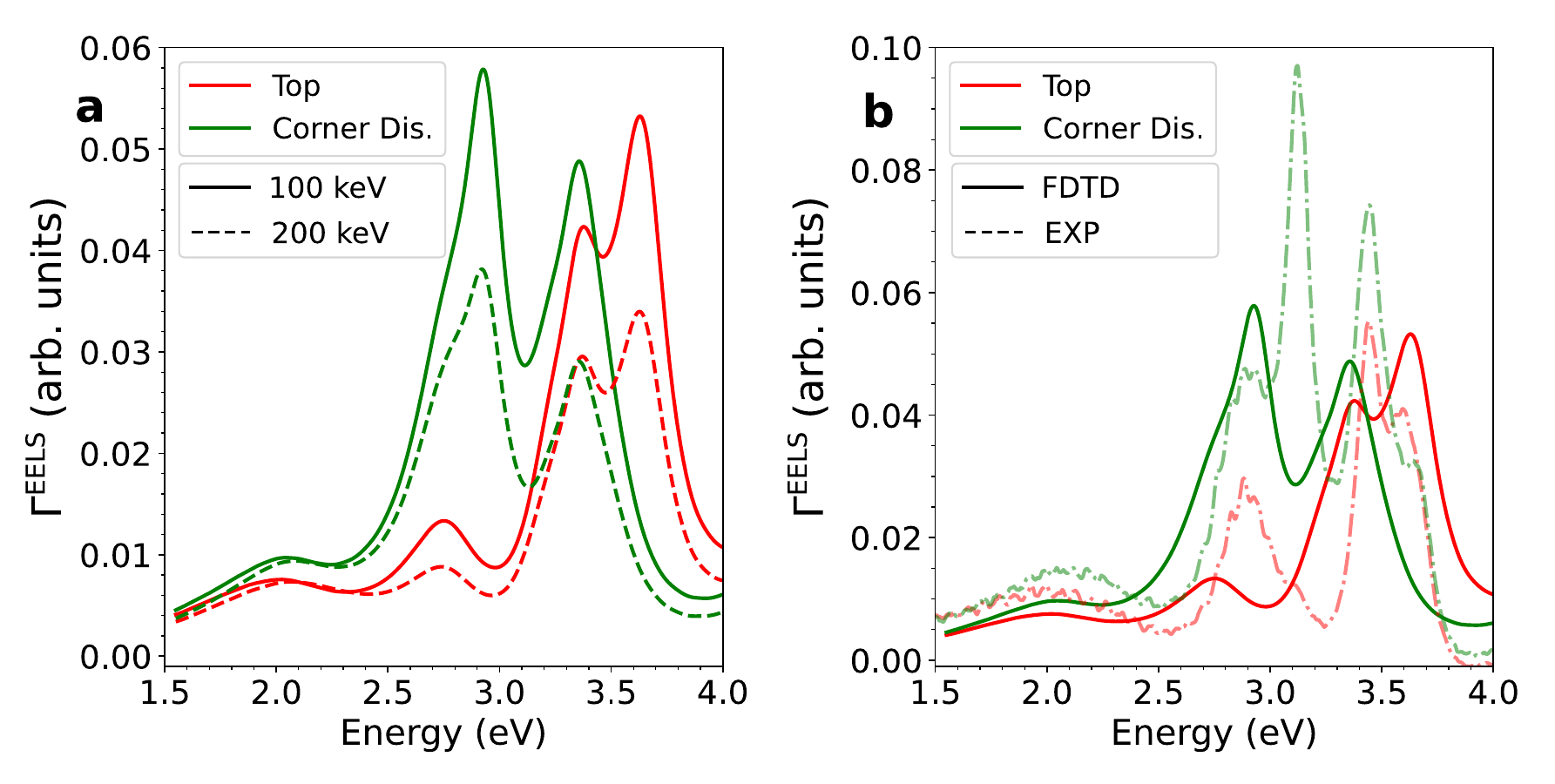}
    \caption{EELS probability comparison for different electron energies and between experiment and simulation. (a) FDTD simulated spectra for 200 keV and 100 keV electrons. (b) Comparison, for 100 keV electrons, between the experimental results and the simulations.}
    \label{fig:SUP_PART02_SPECTRE}
\end{figure}

For the cube-SiO$_2$ substrate system, the FDTD-simulated loss probability was obtained using a 1.25 nm mesh size with electric dipoles evenly spaced by $\sim$ 8 nm. The SiO$_2$ substrate was 500 nm thick and the silver nanocube had 100 nm in side length. Figure \ref{fig:SUP_PART02_SPECTRE}a shows the results for the obtained spectra for the probe at the top edge (opposite from the interface Ag - SiO$_2$) and top corner positions for 200 keV and 100 keV electrons, in which the loss probability changes by less than a relative factor of two (compared to the factor of 100 for the bare resonator). In Figure \ref{fig:SUP_PART02_SPECTRE}b, we compare the simulated and the experimental results. The simulated modes are redshifted with respect to the experimental results in all observed modes (except for the distal face), presumably due to variations in the effective index of refraction between experiment and simulations.

\begin{figure}[H]
    \centering
    \includegraphics[width=0.8\textwidth]{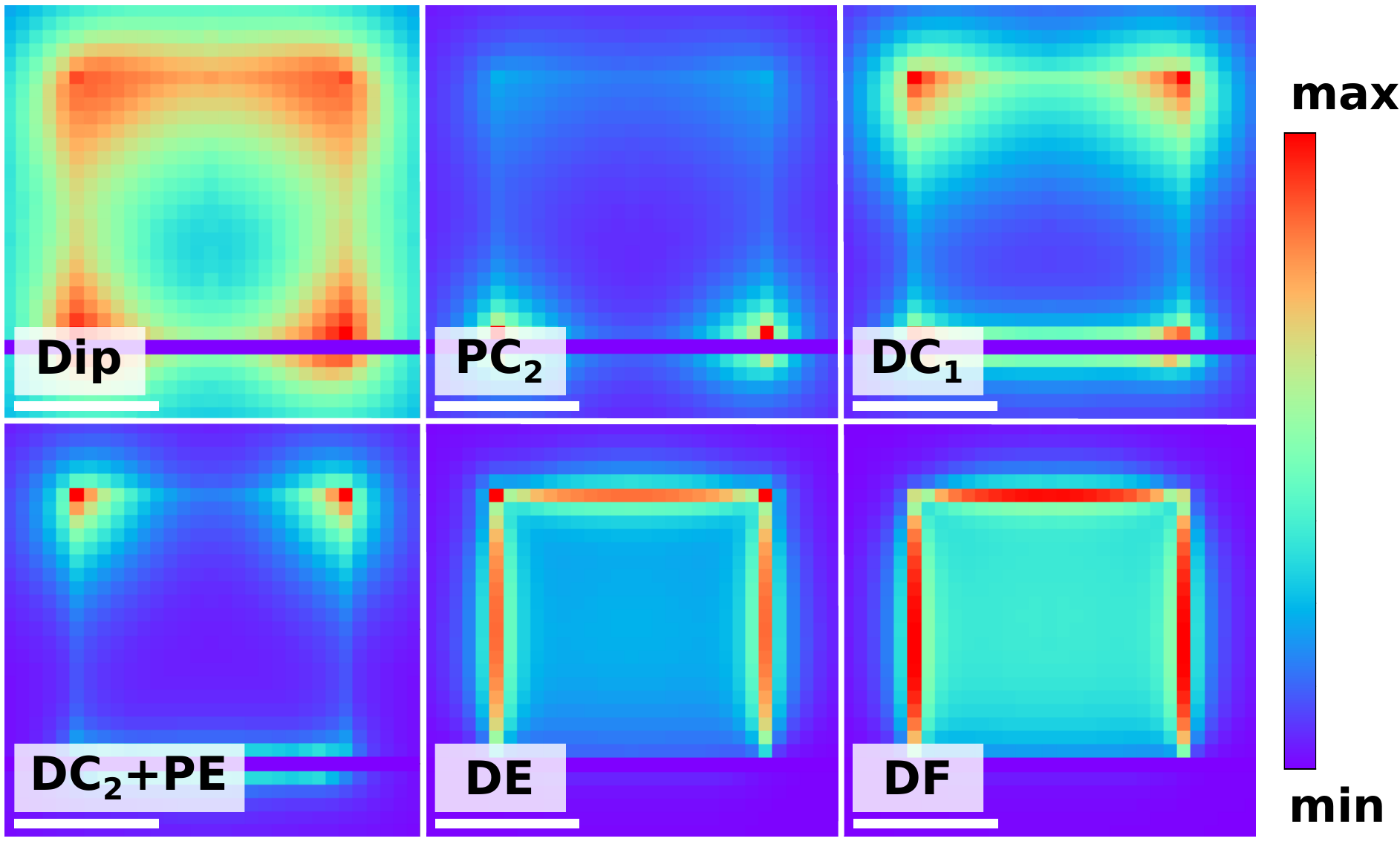}
    \caption{FDTD-simulated spectral maps for the cube-substrate system, including the dipolar (Dip), second proximal corner (PC$_2$), first distal corner (DC$_1$), second distal corner (DC$_2$), proximal edge (PE), distal edge (DE), and distal face (DF) modes. Scale bars are 50 nm.}
    \label{fig:SUP_PART02_FDTD_SPIM}
\end{figure}

Although redshifted, all experimental observed resonances were also observed in FDTD. In order to make a correspondence between them, the absolute value of the near electric field was obtained (shown in Figure 
3c
. This correspondence can also be done through simulating a full spectral image by rastering the electron probe position in a square mesh containing the cube and part of the substrate. For this, the mesh size was increased to 2.50 nm while dipoles were evenly spaced by $\sim$ 25 nm intervals in order to reduce the simulation time. The electron impact position grid was 150 nm wide in both the x-y plane (z being the electron propagation direction) and the impact distance step size was 5 nm. Only minor changes were found in the spectral maps with respect to each single loss spectrum at the corresponding probe positions despite the alleviated simulation conditions, thus pointing to good simulation convergence. Figure \ref{fig:SUP_PART02_FDTD_SPIM} shows the simulated spectral image under the above conditions for the resonances listed in the main text (Figure 
3c),
with a remarkable resemblances between them. Note that simulations have rendered the proximal edge mode (PE) as very faint and close to the second distal corner mode (DC$_2$), similarly to experiment. Figure \ref{fig:SUP_PART02_FDTD_SPIM} for DC$_2$ + PE clearly shows a proximal edge component, although much less strong than the distal corner.

Finally, note that a smaller order proximal order mode PC$_1$ has not been observed in EELS, despite being predicted in the FDTD simulations (Figure
3).
This mode is known to be close to the global dipole and very diffuse, which makes it very challenging to detect experimentally \cite{3DSurfacePlasmon}. It is also important to note that unequivocally characterizing the entire cube-substrate modes requires an extreme degree of electron beam monochromaticity. Experimentally, mode PC$_2$ is centered at 2.813 eV and mode $DC_1$ is centered at 2.898 eV. The 85 meV distance is smaller than the $\sim$ 250 meV mode FWHM, and thus, mode overlap is expected in our spectral maps.

\clearpage
\subsection{D - CL results for higher-order modes in the coupled WGM-plasmon system}

\begin{figure}[H]
    \centering
    \includegraphics[width=0.9\textwidth]{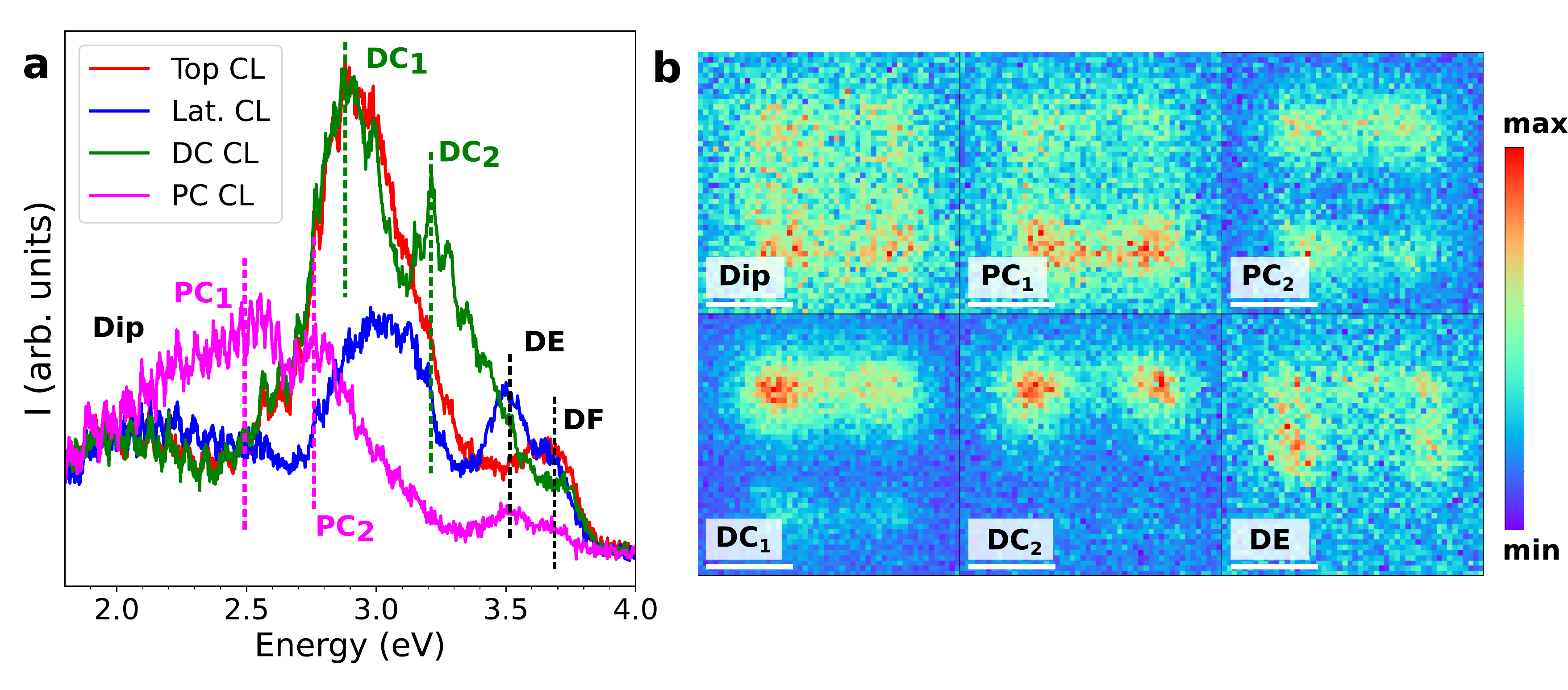}
    \caption{Experimental CL results. (a) CL spectra for the cube-substrate system at four different probe positions. (b) Spectral maps for the dipolar (Dip), first and second proximal corner (PC$_1$ and PC$_2$), first and second distal corner (DC$_1$ and DC$_2$), and distal edge (DE) modes. Scale bars are 50 nm.}
    \label{fig:SUP_D}
\end{figure}

The CL lateral and top probe positions presented in the main text relied on spectra obtained by a spectral image of the cube-SiO$_2$ system. For each of the eigenmodes, CL measures the radiative component of the electromagnetic density of states along the electron propagation direction \cite{LinkEELSCl}. For high-net dipole-moment modes, such as the first dipolar mode (Dip), CL is expected to be intense with respect to the total absorption (performed by EELS in Figure 
3)
. For the higher-order modes, in which the net dipole moment approaches zero, the electron energy absorbed is dissipiated mostly in the form of ohmic losses. However, cube-substrate hybridization and retardation affect the radiation emitted from these modes, as can be seen in Figure \ref{fig:SUP_D}. In this case, they tend to be more probe-position-dependent as the noted deviations are also spatially dependent. We have identified most of the modes revealed by EELS, including the first-order proximal mode PC$_1$ that was difficult to resolve in EELS because of its closeness to the dipolar mode (Dip) and the tail of the much stronger PC$_2$. Additionally, PC$_1$ and PC$_2$ can be better spatially resolved due to the fact that thick regions of the sample can be probed in CL, differently from EELS, in which penetrating trajectories in such a thick sample are very challenging. Note that PC$_2$ radiates less than PC$_1$ due to its higher order (the same happens to DC$_2$ with respect to DC$_1$). Finally, it is worth mentioning that DC$_2$, although with a small radiative contribution, showed coupling to gallery-modes, which reinforces the argument that this coupling is dependent on the net dipole moment. 

\clearpage
\subsection{E - MNP polarizability}

\begin{figure}[H]
    \centering
    \includegraphics[width=0.8\textwidth]{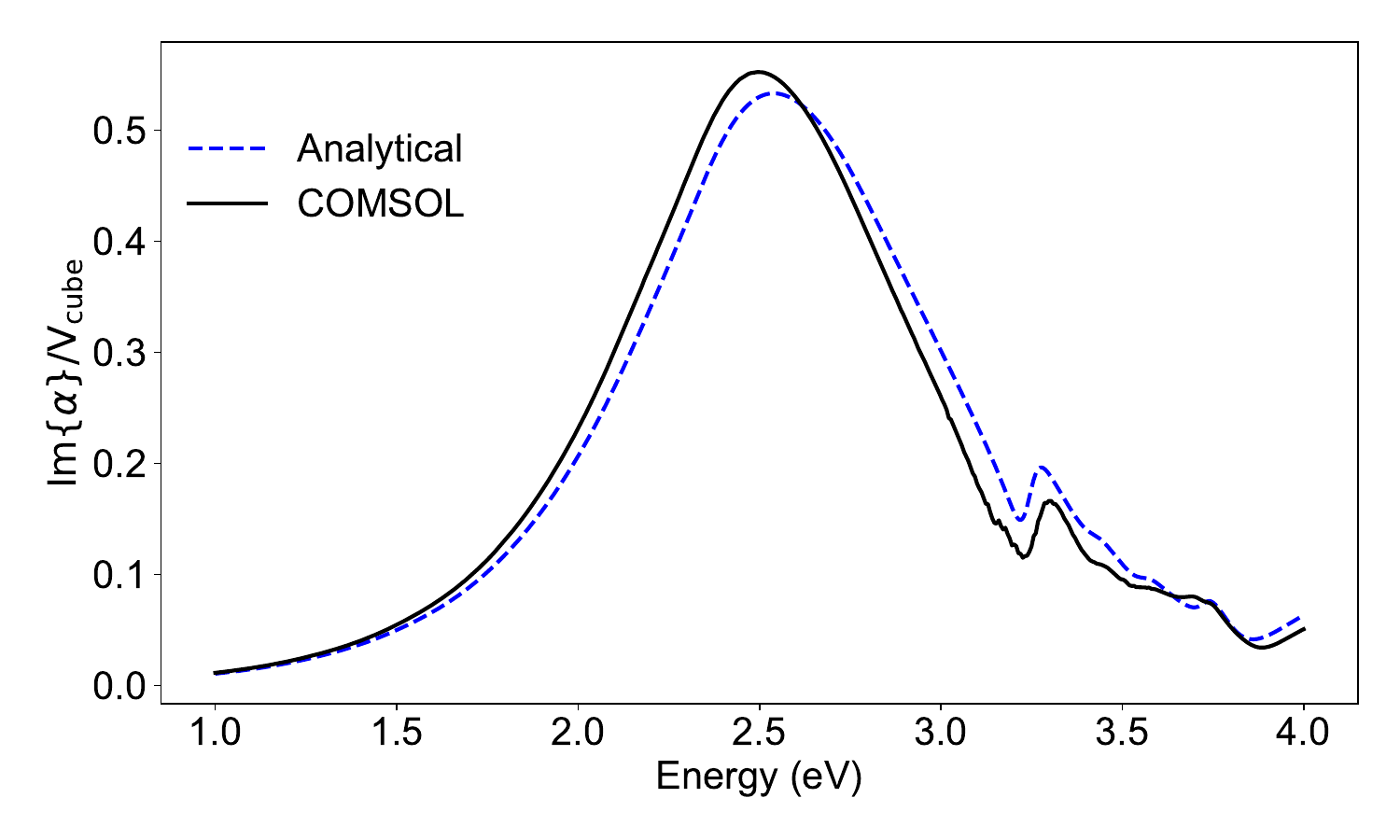}
    \caption{Comparison of the polarization obtained through COMSOL and the analytical form based on the quasistatic approximation with retardation corrections as described in the text.}
    \label{fig:SUP_E_POL}
\end{figure}

In the analytic model, the nanocube is described through as a dipolar point scatterer of isotropic polarizability $\alpha(\omega)$ placed at the particle center. In order to estimate $\alpha(\omega)$, we adopt the model of Fuchs and Sihvola \cite{Fuchs1975, Sihvola2004} for the quasistatic polarizability of the cube,
\begin{equation}
    \alpha_{QS}=\frac{V}{4\pi}\sum_m \frac{C_m}{(\epsilon/\epsilon_h -1)^{-1} + n_m},
    \label{eq: polarisability_QS}
\end{equation}
where $V$ is the cube volume, $\epsilon$ is its dielectric function (taken from tabulated data \cite{JC1972} for silver), $\epsilon_h$ is the host permittivity (1 for vacuum in the present case), $m$ is a mode index, and $C_m$ and $n_m$ are tabulated in Ref. \citenum{Fuchs1975}. We have further obtained the cube polarizabitily using a finite-difference numerical method (COMSOL). Comparison of the quasistatic polarizability and the numerical simulation reveals large discrepencies, which can be attributed to retardation effects. The latter can be incorporated in the analytical model through the precription \cite{Yurkin2007}
\begin{equation}
    \alpha=\frac{\alpha_{QS}}{1-\frac{\alpha_{QS}}{V}M},
    \label{eq: polarisability_QS_corrected}
\end{equation}
where $M=(\frac{4 \pi}{3})^{1/3}(kd)^2+i\frac{2}{3}(kd)^3$, $k=\omega/c$, and $d$ is the cube edge length. This prescription brings the model polarizability in close agreement with the numerical simulation, as shown in Figure \ref{fig:SUP_E_POL}. This analysis confirms the robustness of the polarizability here used for the analytical model of the MNP-sphere system, although of course it does not account for higher-order particle plasmons.

\clearpage
\subsection{F - Additional results from the analytical model}

\begin{figure}[H]
    \centering
    \includegraphics[width=1.0\textwidth]{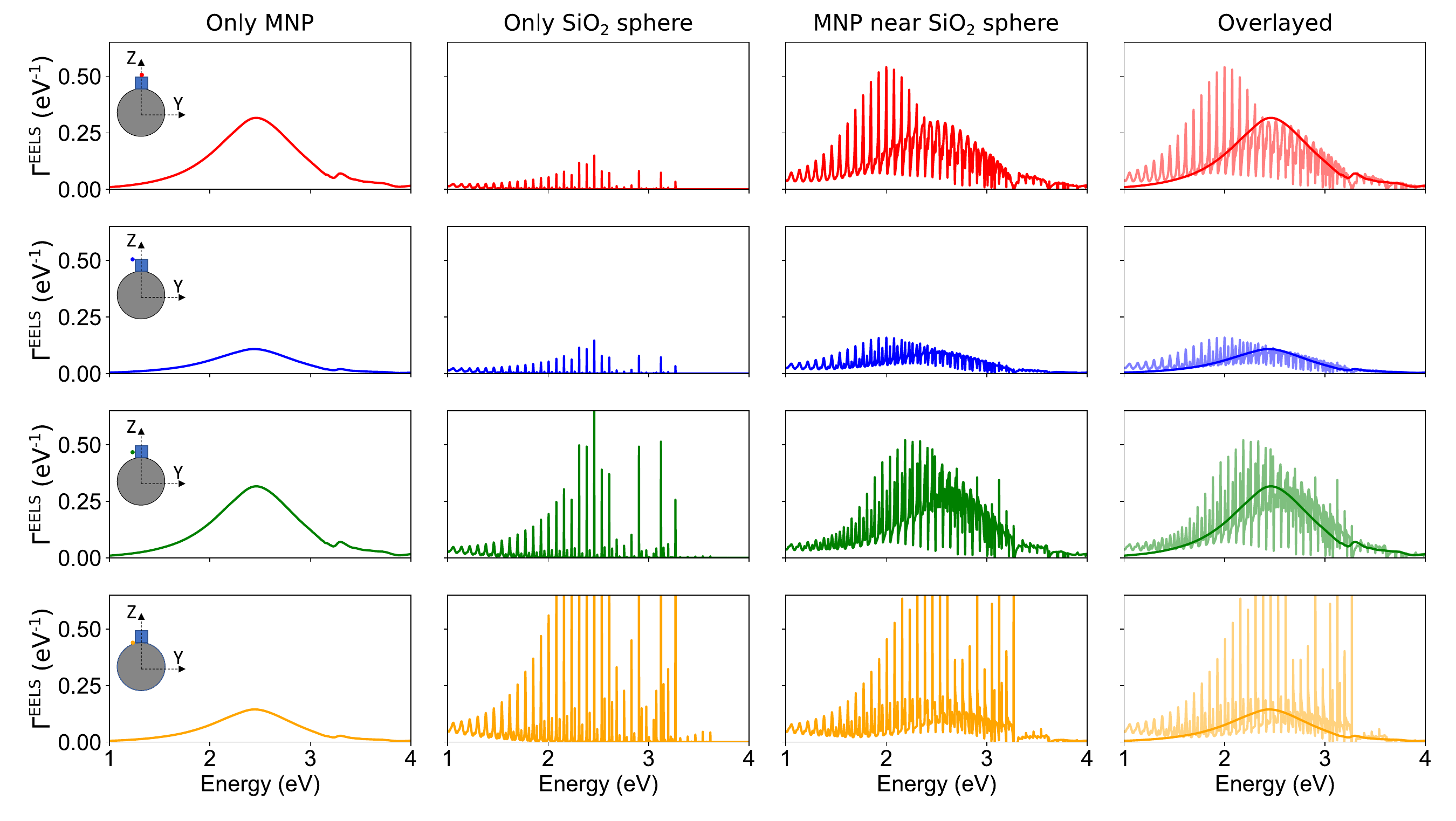}
    \caption{EELS probability calculated by assimilating the MNP to a dipolar scatterer and retaining the full multipolar response of the sphere. We compare spectra for the MNP alone (leftmost column), the sphere alone (second-left column), the hybrid MNP-sphere system (third column), as well as the overlay of only-sphere and MNP+sphere systems spectra (rightmost column) for different positions of the electron beam (see insets in the left column) and an electron energy of 200\,keV.}
    \label{fig:SUP_F_200}
\end{figure}
\begin{figure}[H]
    \centering
    \includegraphics[width=1.0\textwidth]{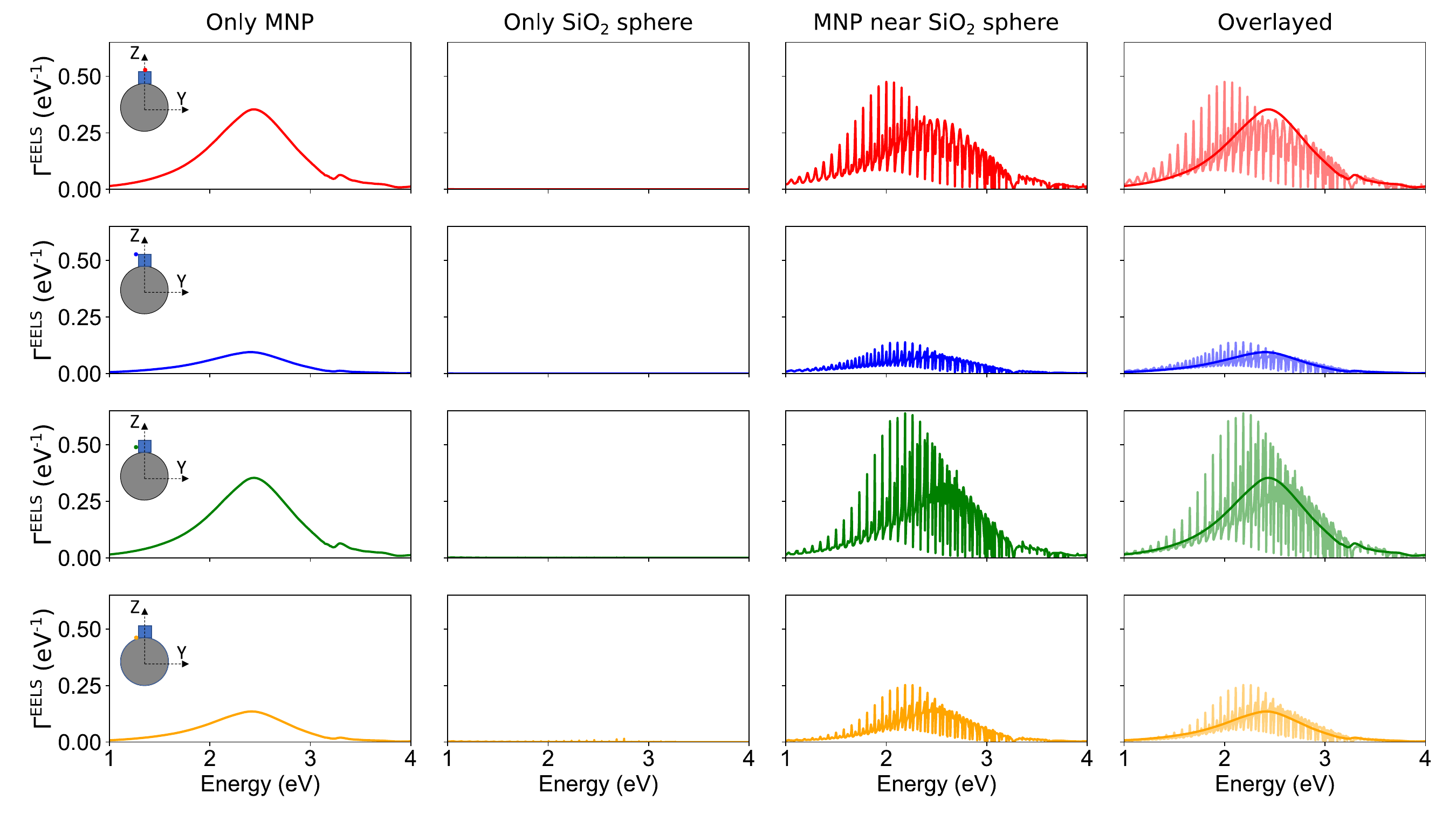}
    \caption{Same as Figure\ \ref{fig:SUP_F_200}, but for 100\,keV electrons.}
    \label{fig:SUP_F_100}
\end{figure}

We show in Figures\ \ref{fig:SUP_F_200} and \ref{fig:SUP_F_100} additional results obtained from the analytical model discussed in the main text for electrons moving with energies of 200\,keV and 100\,keV, respectively.

For 200 keV, Figure\ \ref{fig:SUP_F_200} shows that the coupling is also sensitive to the electron beam position, even in the sphere without a MNP, because the Mie resonances under consideration become more evanescent outside the dielectric for higher orbital angular momentum. For 100\,keV electrons (Figure\ \ref{fig:SUP_F_100}), excitation of Mie resonances in the sphere without the MNP is negligible because of the mismatch between the electron velocity and the orbital group velocity of the whispering-gallery modes; however, the presence of a MNP acts as a mediator in the coupling to Mie modes, which leave sizable signatures in EELS for the hybrid system.

\clearpage